\documentclass[preprint,showpacs,preprintnumbers,amsmath,amssymb]{revtex4}
\usepackage{graphics,epsfig,subfigure}
\usepackage{color}

\begin{document}
\newcommand{\PR}[1]{\ensuremath{\left[#1\right]}} 
\newcommand{\PC}[1]{\ensuremath{\left(#1\right)}} 
\newcommand{\PX}[1]{\ensuremath{\left\lbrace#1\right\rbrace}} 
\newcommand{\BR}[1]{\ensuremath{\left\langle#1\right\vert}} 
\newcommand{\KT}[1]{\ensuremath{\left\vert#1\right\rangle}} 
\newcommand{\MD}[1]{\ensuremath{\left\vert#1\right\vert}} 

\title{Thick braneworld model in nonmetricity formulation of general relativity and its stability}
\author{Qi-Ming Fu$^{1,}$\footnote{fuqiming@snut.edu.cn},
        Li Zhao$^{2,}$\footnote{lizhao@lzu.edu.cn},
        and Qun-Ying Xie$^{2,3,}$\footnote{xieqy@lzu.edu.cn, corresponding author}}

\affiliation{$^{1}$Institute of Physics, Shaanxi University of Technology, Hanzhong 723000, China \\
             $^{2}$Institute of Theoretical Physics $\&$ Research Center of Gravitation, Lanzhou University, Lanzhou 730000, China \\
             $^{3}$School of Information Science and Engineering, Lanzhou University, Lanzhou 730000, China}

\begin{abstract}
  In this paper, we study the thick brane system in the so-called $f(Q)$ gravity, where the gravitational interaction was encoded by the nonmetricity $Q$ like scalar curvature $R$ in general relativity. With a special choice of $f(Q)=Q-b Q^n$, we find that the thick brane system can be solved analytically with the first-order formalism, where the complicated second-order differential equation is transformed to several first-order differential equations. Moreover, the stability of the thick brane system under tensor perturbation is also investigated. It is shown that the tachyonic states are absent and the graviton zero mode can be localized on the brane. Thus, the four-dimensional Newtonian potential can be recovered at low energy. Besides, the corrections of the massive graviton Kaluza-Klein modes to the Newtonian potential are also analyzed briefly.
\end{abstract}





\maketitle

\section{Introduction}

General relativity (GR) has been successfully tested for many years. However, it
breaks down at quantum level or galactic scale for its nonrenormalization and disability of explaining dark matter and dark energy. Thus, the theory of GR needs to be revised at high energy and galactic scale. There are many extended theories of gravity in literatures. Most of them are based on the metric-compatible and torsionless Levi-Civita connection. In this paper, we are interested in the so-called symmetric teleparallel equivalent of general relativity (STEGR) \cite{Nester1999}, where the gravitational interaction is manifested by the nonmetricity $Q$, and the curvature and torsion of the spacetime are vanishing.
A straightforward extension of the original STEGR is $f(Q)$ gravity introduced in Ref.~\cite{Jimenez2018a}, where the authors introduced a simpler geometrical formulation of GR with vanishing affine connection, i.e., the coincident gauge, and the spacetime described by this theory is trivially connected.

As a novel modified gravitational theory, many investigations on $f(Q)$ gravity and its extensions have been done in different contexts.
For instance, a set of constraints on $f(Q)$ gravity by observational data were explored in Ref.~\cite{Lazkoz2019}, where the $f(Q)$ Lagrangian was transferred to a function of the redshift. In Ref.~\cite{Mandal2020a}, the authors investigated the energy conditions for some explicit $f(Q)$ models, which can be used to fix some free parameters and give some restrictions on the form of $f(Q)$. In Ref.~\cite{Jianbo2019}, the authors studied the acceleration of the cosmic expansion in $f(Q)$ gravity and found that the density and pressure of the dark energy can be expressed as a function of geometry, which indicates a geometric dynamical dark energy model. The propagation of the gravitational wave around Minkowski spacetime in a general class of STEGR was studied in Refs.~\cite{Hohmann2019,Soudi2019}, where the authors focused on its velocity and polarizations. Some extensions of $f(Q)$ gravity by coupling the nonmetricity $Q$ to scalar field, the trace of the energy-momentum tensor, and the matter Lagrangian were considered in Refs.~\cite{Jarv2018,Runkla2018,Harko2018,Yixin2019}.
Besides, there are other important investigations on $f(Q)$ gravity and one can see Refs.~\cite{Jimenez2018b,Jimenez2019,Dialektopoulos2019,Jimenez2020,Mandal2020b} for uncompleted lists.

On the other hand, the braneworld theory has been extensively investigated for many years, which considers our four-dimensional world just a brane embedded in a higher-dimensional spacetime.
According to the thickness of the brane, the braneworld models can be mainly divided into two kinds, i.e., the thin braneworld models and thick braneworld models.
The most investigated thin braneworld models are the Randall-Sundrum braneworld models \cite{rs1,rs2} and their extensions \cite{Davoudiasl2000,Gherghetta2001,Huber2001}. In the thin braneworld models, the gravity and different matter fields are localized at different locations along the extra dimension, which can be used to explain not only the gauge hierarchy problem but also the fermion mass hierarchy problem. Besides, the thin braneworld models also provide us alternative approaches to address the cosmological constant problem, the nature of dark energy and dark matter. However, because of the vanishing thickness of the brane, the thin braneworld models only can be treated as an effective theory of a more fundamental theory, where any objects would have a minimal length scale. Thus, it is more realistic to investigate the thick braneworld models, where the brane is generated by one or more background scalar fields and the energy density of the brane is a smooth function of the extra dimension \cite{DeWolfe2000,Csaki2000,Gremm2000,Giovannini2001,Minamitsuji2006,Dzhunushaliev2008,Liu2011,Zhong2011,Bazeia2015,Dzhunushaliev2020,Bazeia2020a,Bazeia2020b,Sui2020,Cui2020,Moreira2101} instead of a Dirac delta function in the thin braneworld models. In the thick braneworld models, although the size of the extra dimension is infinity, the matter fields corresponding to the standard model are confined on the brane \cite{Melfo2008,Liu2008,Liu2009,Dantas2015,Vaquera-Araujo2015,Arai2013,Zhao2015} and the localized graviton zero mode produces the four-dimensional Newtonian potential \cite{Kobayashi2002,Andrianov2008,Barbosa-Cendejas2008,Andrianov2013,Barbosa-Cendejas2005,Veras2016}, which are two key points reconstructing our effective four-dimensional world. One can see Ref.~\cite{Liu2017} for a brief review.

Although there are many investigations on $f(Q)$ gravity, the thick braneworld model in this theory has not been considered yet. As a novel modified gravitational theory, it is interesting to know whether one can construct the thick braneworld model in this theory and what the effects are of the nonmetricity on the thick brane system and its stability under tensor perturbation.

This paper is organised as follows: In Sec.~\ref{fieldeq}, we give a brief review of $f(Q)$ gravity and then the thick brane system is solved analytically with the first-order formalism. In Sec.~\ref{tensorper}, the stability of the thick brane system under tensor perturbation and the localization of the graviton zero mode are investigated. Section \ref{conclusion} comes with the conclusion.

\section{Thick brane in $f(Q)$ gravity}~\label{fieldeq}

In metric-affine geometry, the metric and affine connection are treated as two independent objects. In this framework, the metric encodes distances and angles, while the affine connection defines the covariant derivatives and parallel transport. As known from differential geometry, the general affine connection can always be decomposed as
\begin{eqnarray}
\Gamma^{H}_{~MN}=\hat{\Gamma}^{H}_{~MN}+K^{H}_{~MN}+L^{H}_{~MN},
\end{eqnarray}
where Roman indices denote spacetime coordinates, i.e., $H,I,M,N, \ldots =0,1,2,3,5$, and the Levi-Civita connection is
\begin{eqnarray}
\hat{\Gamma}^{H}_{~MN}\equiv \frac{1}{2}g^{HI}(\partial_{M}g_{IN}+\partial_{N}g_{IM}-\partial_{I}g_{MN}).
\end{eqnarray}
The second term is the contortion:
\begin{eqnarray}
K^{H}_{~MN}\equiv \frac{1}{2}g^{HI}(T_{M I N}+T_{N I M}+T_{I M N}),
\end{eqnarray}
with the torsion tensor denoted by $T^{H}_{~MN}\equiv \Gamma^{H}_{~MN}-\Gamma^H_{~NM}$. The third term is the disformation tensor expressed as
\begin{eqnarray}
L^{H}_{~MN}\equiv \frac{1}{2}g^{HI}(-Q_{M I N}-Q_{N I M}+Q_{I M N}),
\end{eqnarray}
where the nonmetricity tensor is defined by
\begin{eqnarray}
Q_{HMN}\equiv \nabla_{H}g_{MN}=\partial_{H}g_{MN}-\Gamma^{I}_{~H M}g_{I N}-\Gamma^{I}_{~H N}g_{M I},
\end{eqnarray}
endowed with two independent traces
\begin{eqnarray}
Q_{M}\equiv Q_{M~N}^{~~N}, \quad\quad \tilde{Q}_{M}\equiv Q^{N}_{~MN}.
\end{eqnarray}
Besides, it is useful to introduce the nonmetricity conjugate
\begin{eqnarray}~\label{P}
P^K_{~MN}=-\frac{1}{4}\Big(Q^K_{~MN}-2Q_{(M~N)}^{~~~K}-Q^K g_{MN}+\tilde{Q}^K g_{MN}+\delta^K_{(M}Q_{N)}\Big).
\end{eqnarray}

Taking the torsion and curvature to be vanishing, one obtain the so-called symmetric teleparallel equivalent of general relativity (STEGR) with the Lagrangian \cite{Nester1999}
\begin{eqnarray}
\mathcal{L}=\frac{1}{2}\sqrt{-g}Q,
\end{eqnarray}
where the nonmetricity scalar is defined by $Q=Q_{HMN}P^{HMN}$. Since the nonmetricity scalar differs from the scalar curvature only by  a boundary term, STEGR is equivalent to general relativity. However, this equivalence is not preserved in their extensions, i.e., $f(Q)$ and $f(R)$, which lead to two quite different gravitational field equations.
In this paper, we are interested in the thick braneworld model in five-dimensional $f(Q)$ gravity. The action is read as \cite{Jimenez2018a}
\begin{eqnarray}
S=\int d^5x\sqrt{-g} \left[\frac{1}{2\kappa}f(Q)-\frac{1}{2}\partial_M\phi\partial^M\phi-V(\phi)\right], ~\label{actionf}
\end{eqnarray}
where $\kappa=8\pi G_5$ with $G_5$ the five-dimensional Newtonian gravitational constant.

Taking the variation of the action (\ref{actionf}) with respect to the metric $g_{MN}$, the scalar field $\phi$, and the connection $\Gamma^K_{~MN}$, one can obtain the equations of motion
\begin{eqnarray}
\frac{2}{\sqrt{-g}}\nabla_{K}\big(\sqrt{-g}f_Q P^K_{~~MN}\big)-\frac{1}{2}g_{MN}f+f_Q\big(P_{MKL}Q_N^{~~KL}-2Q_{KM}^{~~~~L}P^K_{~~NL}\big)&=&\kappa T_{MN},~\label{EoMofEinstein} \\
\frac{1}{\sqrt{-g}}\nabla_M\big(\sqrt{-g}\nabla^M\phi\big)-V_{\phi}&=&0,~\label{seom} \\
\nabla_M\nabla_N\big(\sqrt{-g}f_Q P^{MN}_{~~~~K}\big)&=&0,~\label{ceom}
\end{eqnarray}
where $f_Q\equiv\frac{df}{dQ}$, $V_{\phi}\equiv\frac{dV}{d\phi}$, and the energy-momentum tensor is
\begin{eqnarray}
T_{MN}=\partial_M\phi\partial_N\phi-\frac{1}{2}g_{MN}\partial_K\phi\partial^K\phi-g_{MN}V(\phi).~\label{energytensor}
\end{eqnarray}

In general, the metric for a static flat thick brane can be assumed as
\begin{eqnarray}
ds^2=\text{e}^{2A(y)}\eta_{\mu\nu}dx^{\mu}dx^{\nu}+dy^2,~\label{metric}
\end{eqnarray}
with $\text{e}^{2A(y)}$ the so-called warp factor and $y$ the extra-dimensional coordinate. The Greek indices $\mu,\nu,\ldots$ run from 0 to 3. In the coincident gauge with $\Gamma^H_{~MN}=0$, the covariant derivative reduces to ordinary derivative. Then, Eqs.~(\ref{EoMofEinstein}) and (\ref{seom}) for the thick brane system reduce to
\begin{eqnarray}
-6 A' f_Q'-6 f_Q \left(A''+4 A'^2\right)+f-2\kappa V-\kappa\phi '^2&=&0,~\label{eq1} \\
12 f_Q A'^2-\frac{f}{2}+\kappa V-\frac{1}{2}\kappa \phi '^2&=&0,~\label{eq2} \\
\phi''+4A'\phi'-V_{\phi}&=&0,
\end{eqnarray}
where the prime denotes the derivative with respect to $y$. Besides, it can be easily shown that Eq.~(\ref{ceom}) can always be satisfied with the metric ansatz (\ref{metric}).
Focusing on an explicit choice $f(Q)=Q-b^2 Q^n$ with $Q=12A'^2$, and from Eqs.~(\ref{eq1}) and (\ref{eq2}), we get
\begin{eqnarray}
6 A'' \left(1 -12^{n-1} n b^2 (2 n-1) A'^{2n-2}\right)+2\kappa \phi '^2=0.~\label{supereq}
\end{eqnarray}

In the following, we will show that the above second-order differential equation can be solved analytically with the first-order formalism. By plugging the assumption
\begin{eqnarray}
A'=-\kappa W(\phi),
\end{eqnarray}
into Eq.~(\ref{supereq}), we immediately obtain
\begin{eqnarray}
\phi'=3 W_{\phi} \left(1-12^{n-1}b^2 n (2 n-1) (\kappa W)^{2 n-2}\right)~\label{eqphi}.
\end{eqnarray}
Then, from Eq.~(\ref{eq1}), the potential for the background scalar field can be solved as
\begin{eqnarray}
V&=&\frac{12^n b^2 (2 n-1) (\kappa W)^{2 n-2}}{8} \left(4\kappa W^2-3n W_{\phi }^2 \left( 12^{n-1} b^2 n (1-2 n) (\kappa W)^{2 n-2}+1\right)\right) \nonumber\\
 &+&\frac{9}{2} W_{\phi }^2 \left( 12^{n-1} b^2 n (1-2 n) (\kappa W)^{2 n-2}+1\right)-6\kappa W^2.
\end{eqnarray}

Now, the solutions of the thick brane system are completely determined by the so-called superpotential function $W(\phi)$ required to be specified. We will take two explicit $W(\phi)$ for examples and present the corresponding thick brane solutions. The first example is $W(\phi)=\frac{k\phi_0^2}{3} \sin( \phi/\phi_0)$ and $n=\frac{1}{2}$. It is obvious that the second term in parentheses on the right-hand side of Eq.~(\ref{eqphi}) vanishes with this special $n$, and the thick brane system can be easily solved as
\begin{eqnarray}
A(y)&=&\frac{1}{3}\kappa\phi_0^2\ln[\text{sech}(ky)], \\
\phi(y)&=&\phi_0 \arcsin[\tanh (k y)], \\
V(\phi)&=&\frac{1}{12} k^2 \phi _0^2 \left[\left(4 \kappa  \phi _0^2+3\right) \cos(2\phi/\phi_0)-4 \kappa  \phi _0^2+3\right].
\end{eqnarray}

The second example is a linear function $W(\phi)=k\phi$ and $n=2$. The solutions for this brane system are solved as
\begin{eqnarray}
A(y)&=&\frac{3\kappa}{c^2 k^2}\ln \left[\text{sech}\left(c k^2 y\right)\right],~\label{solA2} \\
\phi(y)&=&\frac{3}{c k}\tanh \left(c k^2 y\right), \\
V(\phi)&=&\frac{1}{18} k^2 \left(\phi ^2 \left(c^2 k^2+6 \kappa \right) \left(c^2 k^2 \phi ^2-18\right)+81\right),
\end{eqnarray}
where we have introduced a new parameter $c\equiv 18\sqrt{2}b\kappa$.

The energy density of the thick brane is defined as $\rho\equiv T_{MN}u^Mu^N-V_0$, where $u^M$ denotes the velocity of a static observer and $V_0$ stands for the scalar vacuum energy density. Then,
the energy density for the above two brane systems can be expressed as
\begin{eqnarray}
\rho=\frac{1}{3} k^2 \phi _0^2 \left(\kappa  \phi _0^2+3\right) \text{sech}^2(k y), \quad (n=1/2)
\end{eqnarray}
and
\begin{eqnarray}
\rho=9 \left(k^2+\frac{3\kappa}{c^2}\right) \text{sech}^4\left(c k^2 y\right). \quad (n=2)
\end{eqnarray}

\begin{figure*}[htb]
\begin{center}
\subfigure[$\text{e}^{2A(y)}$]  {\label{A}
\includegraphics[width=6cm]{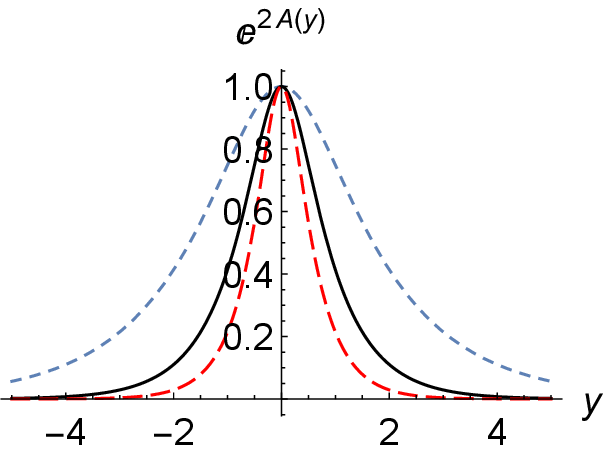}}
\subfigure[$\phi(y)$]  {\label{phi}
\includegraphics[width=6cm]{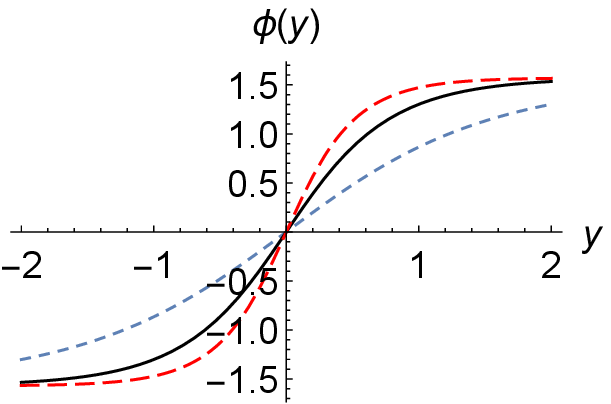}} \\
\subfigure[$V(\phi)$]  {\label{V}
\includegraphics[width=6cm]{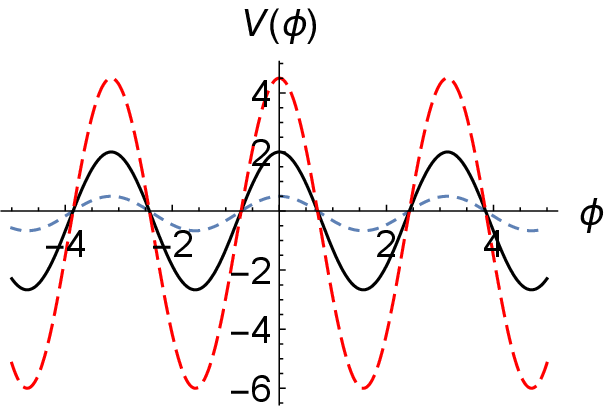}}
\subfigure[$\rho(y)$]  {\label{rho}
\includegraphics[width=6cm]{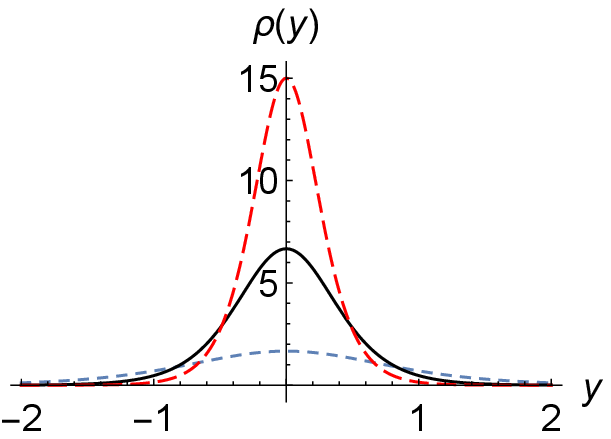}}
\end{center}
\caption{The shapes of the warp factor $\text{e}^{2A}$, background scalar field $\phi(y)$, scalar potential $V(\phi)$, and energy density $\rho(y)$ for the first brane solutions. The parameters are set to $\kappa=\phi_0=1$, $k=1$ for blue short dashed lines, $k=2$ for black lines, and $k=3$ for red long dashed lines.}
\label{brane1}
\end{figure*}

\begin{figure*}[htb]
\begin{center}
\subfigure[$\text{e}^{2A(y)}$]  {\label{A2}
\includegraphics[width=6cm]{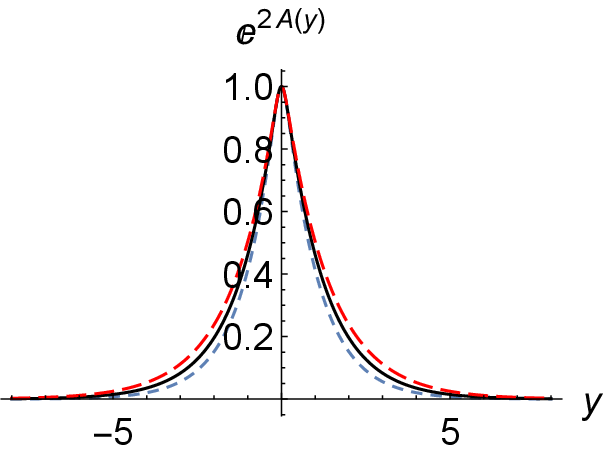}}
\subfigure[$\phi(y)$]  {\label{phi2}
\includegraphics[width=6cm]{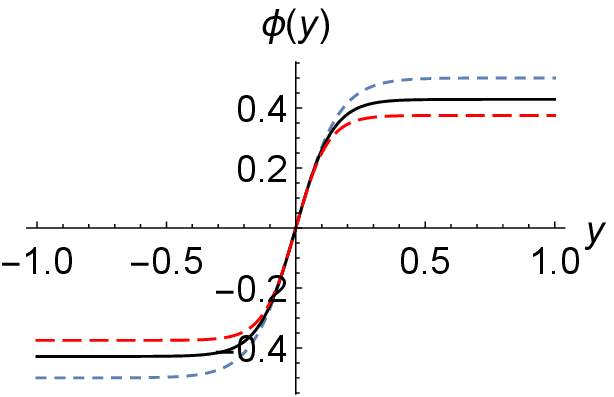}} \\
\subfigure[$V(\phi)$]  {\label{V2}
\includegraphics[width=6cm]{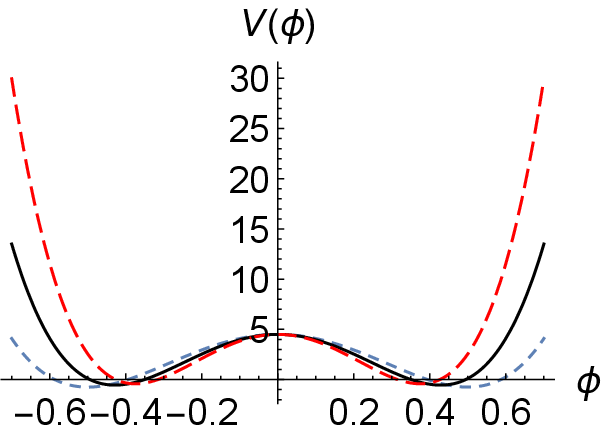}}
\subfigure[$\rho(y)$]  {\label{rho2}
\includegraphics[width=6cm]{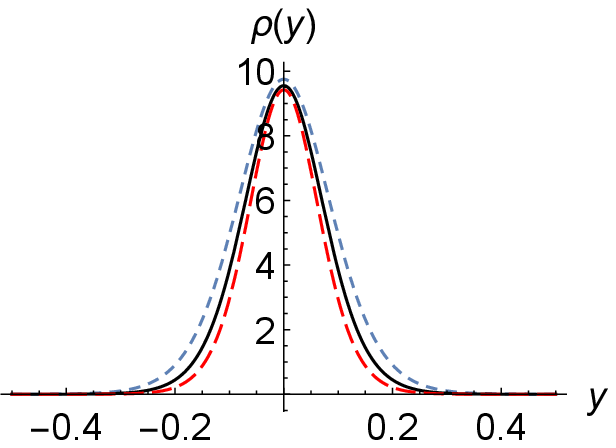}}
\end{center}
\caption{The shapes of the warp factor $\text{e}^{2A}$, background scalar field $\phi(y)$, scalar potential $V(\phi)$, and energy density $\rho(y)$ for the second brane solutions. The parameters are set to $k=\kappa=1$, $c=6$ for blue short dashed lines, $c=7$ for black lines, and $c=8$ for red long dashed lines.}
\label{brane2}
\end{figure*}

Figures \ref{brane1} and \ref{brane2} show the shapes of the above two explicit solutions of the thick brane system in $f(Q)$ gravity. For the first example, the parameter $b$ does not affect the brane system since the vanishing of the second term in parentheses on the right-hand side of Eq.~(\ref{eqphi}). For the second example, the parameter $b$ affects the warp factor, the scalar field and the energy density through the parameter $c$. To be explicit, the warp factor becomes wider while the energy density becomes narrower and smaller with increasing $c$, and the amplitude of the scalar field at $y\rightarrow\pm\infty$ decreases with $c$. Besides, the energy density for the thick brane always peaks at $y=0$ for both models, which means a single brane with no inner structure. From the asymptotic behaviors of the warp factor $A(y\rightarrow\pm\infty)\rightarrow -\frac{1}{3}\kappa k\phi_0^2|y|$ for $n=1/2$ and $A(y\rightarrow\pm\infty)\rightarrow -\frac{3\kappa}{c}|y|$ for $n=2$, one can conclude that the spacetimes for both brane systems are asymptotic anti-de Sitter along the fifth dimension.

\section{Tensor Perturbation}~\label{tensorper}

In this section, we will investigate the tensor perturbation of the thick brane system described in the last section. In general, the tensor, vector and scalar perturbations are decoupled from each other. Thus, we can investigate them individually. The metric of the thick brane system under the tensor perturbation is given by
\begin{eqnarray}
ds^2=\text{e}^{2A(y)}(\eta_{\mu\nu}+h_{\mu\nu})dx^{\mu}dx^{\nu}+dy^2. \label{flumetric}
\end{eqnarray}

It should be stressed that we still take the coincident gauge, i.e., $\Gamma^K_{~MN}=0$, even at the perturbation level. Besides, since the tensor perturbation is gauge-invariant, we can adopt the transverse-traceless (TT) conditions, i.e., $\partial^{\mu}h_{\mu\nu}=\eta^{\mu\nu}h_{\mu\nu}=0$. The nonvanishing components of the perturbed nonmetricity tensor are
\begin{eqnarray}~\label{delQ}
\delta Q^{\rho}_{~\mu\nu}&=&\partial^{\rho}h_{\mu\nu}, \nonumber\\
\delta Q^5_{~\mu\nu}&=&2A'\text{e}^{2A}h_{\mu\nu}+\text{e}^{2A}\partial_5 h_{\mu\nu},
\end{eqnarray}
and the perturbed traces are
\begin{eqnarray}~\label{delracQ}
\delta Q_{\mu}&=&\partial_{\mu}h, \quad\quad~~ \delta Q_5=\partial_5 h, \nonumber\\
\delta \tilde{Q}_{\mu}&=&\partial^{\nu}h_{\mu\nu}, \quad\quad \delta \tilde{Q}_5=0,
\end{eqnarray}
where $h\equiv \eta^{\mu\nu}h_{\mu\nu}$. Besides, the perturbed nonmetricity scalar is $\delta Q=3A'\partial_5 h$.

Inserting the perturbed nonmetricity tensor (\ref{delQ}) and their perturbed traces (\ref{delracQ}) into Eq.~(\ref{P}), the perturbed nonmetricity conjugate can be derived as
\begin{eqnarray}
\delta P^{\rho}_{~\mu\nu}&=&-\frac{1}{4}\bigg[\partial^{\rho}h_{\mu\nu}-(\partial_{\mu}h^{\rho}_{~\nu}+\partial_{\nu}h^{\rho}_{~\mu})+\eta_{\mu\nu}(\partial_{\sigma}h^{\sigma\rho}-\partial^{\rho}h)+\frac{1}{2}(\delta^{\rho}_{\mu}\partial_{\nu}h+\delta^{\rho}_{\nu}\partial_{\mu}h)\bigg], \\
\delta P^5_{~\mu\nu}&=&\frac{1}{4}(6A'\text{e}^{2A}h_{\mu\nu}-\text{e}^{2A}\partial_5 h_{\mu\nu}+\text{e}^{2A}\eta_{\mu\nu}\partial_5 h), \\
\delta P^{\rho}_{~5\nu}&=&\delta P^{\rho}_{~\nu 5}=\frac{1}{4}\left(\partial_5 h^{\rho}_{~\nu}-\frac{1}{2}\delta^{\rho}_{\nu}\partial_5 h\right), \\
\delta P^5_{~5\nu}&=&\delta P^5_{~\nu 5}=-\frac{1}{8}\partial_{\nu}h, \\
\delta P^{\rho}_{~55}&=&\frac{1}{4}(\text{e}^{-2A}\partial^{\rho}h+\text{e}^{-2A}\partial_{\sigma}h^{\sigma\rho}), \\
\delta P^5_{~55}&=&0.
\end{eqnarray}

With the expression of the perturbed scalar field $\phi=\bar{\phi}+\delta \phi$, one can get the perturbation of the energy-momentum tensor:
\begin{eqnarray}
\delta T_{\mu\nu}&=&-\text{e}^{2A}\bigg(\frac{1}{2}\bar{\phi}'^2 h_{\mu\nu}+\bar{\phi}'\delta\phi'\eta_{\mu\nu}+V h_{\mu\nu}+V_{\phi}\delta\phi\eta_{\mu\nu}\bigg), \\
\delta T_{5\mu}&=&\bar{\phi}'\partial_{\mu}\delta\phi, \\
\delta T_{55}&=&\bar{\phi}'\delta\phi'-V_{\phi}\delta\phi,
\end{eqnarray}
where $\bar{\phi}=\bar{\phi}(x^{\mu})$ and $\delta\phi=\delta\phi(x^{\mu},y)$ stands for the background scalar field and its perturbation, respectively. Then, inserting the above perturbed quantities into Eq.~(\ref{EoMofEinstein}) and considering the TT conditions, one can obtain the equations of motion for the tensor perturbation:
\begin{eqnarray}
h''_{\mu\nu}+\left(4A'+\frac{f_Q'}{f_Q}\right)h'_{\mu\nu}+\text{e}^{-2A}\Box^{(4)}h_{\mu\nu}=0, \label{fluy}
\end{eqnarray}
where $\Box^{(4)}\equiv \eta^{\mu\nu}\partial_{\mu}\partial_{\nu}$. Under the coordinate transformation $dy=\text{e}^{A}dz$, Eq.~(\ref{fluy}) can be rewritten as
\begin{eqnarray}
\partial_z^2h_{\mu\nu}+\left(3\partial_z A+\frac{\partial_z f_Q}{f_Q}\right)\partial_z h_{\mu\nu}+\Box^{(4)}h_{\mu\nu}=0. \label{fluz}
\end{eqnarray}

Considering the decomposition $h_{\mu\nu}(x,z)=(\text{e}^{-3A/2}f_Q^{-1/2})\varepsilon_{\mu\nu}(x)\text{e}^{-ipx}\psi(z)$ with $p^2=-m^2$, Eq.~(\ref{fluz}) can be transformed into a Schr$\ddot{\text{o}}$dinger-like equation:
\begin{eqnarray}
(-\partial_z^2+U(z))\psi(z)=m^2\psi(z),~\label{scheq}
\end{eqnarray}
where the effective potential $U(z)$ is given by
\begin{eqnarray}
U(z)=\frac{9}{4}(\partial_z A)^2+\frac{3}{2}\partial_z^2 A+\frac{3\partial_zA \partial_z f_Q}{2f_Q}-\frac{1}{4}\left(\frac{\partial_z f_Q}{f_Q}\right)^2+\frac{\partial_z^2f_Q}{2f_Q}.~\label{effpotential}
\end{eqnarray}
Equation (\ref{scheq}) can be easily factorized as
\begin{eqnarray}
\left(\partial_z+\frac{\Xi}{2}\right)\left(-\partial_z+\frac{\Xi}{2}\right)\psi(z)=m^2 \psi(z),~\label{flusy}
\end{eqnarray}
with $\Xi\equiv \left(3\partial_z A+\frac{\partial_z f_Q}{f_Q}\right)$, which indicates that there is no tachyonic states with $m^2\leq 0$, i.e., the brane is stable under the tensor perturbation.

The general solution for the graviton zero mode can be solved as \cite{Cui2020}
\begin{eqnarray}
\psi_0=\text{e}^{3A/2}f^{1/2}_Q\left(C_1+C_2\int\frac{1}{\text{e}^{3A}f_Q}dz\right),~\label{zeromode}
\end{eqnarray}
with $C_1$ and $C_2$ the integration constants. For simplicity, we take the Neumann boundary condition, i.e., $\partial_z\big(\text{e}^{-3A/2}f^{-1/2}_Q\psi_0\big)\big|_{z\rightarrow\pm\infty}=0$. Then, the graviton zero mode reduces to $\psi_0=C_1\text{e}^{3A/2}f^{1/2}_Q$, where the integration constant $C_1$ is determined by the normalization condition $\int \psi_0^2 dz=1$.

For the first thick brane solution with $n=\frac{1}{2}$, since the wave function of the graviton zero mode is imaginary and divergent near the origin of the extra dimension, one can not obtain a localized graviton zero mode in this brane system.

\begin{figure*}[htb]
\begin{center}
\subfigure[$U(z)$]  {\label{U}
\includegraphics[width=6cm]{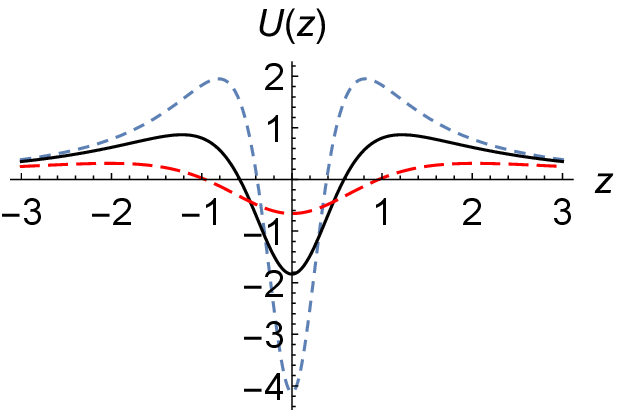}}
\subfigure[$\psi_0(z)$]  {\label{psi}
\includegraphics[width=6cm]{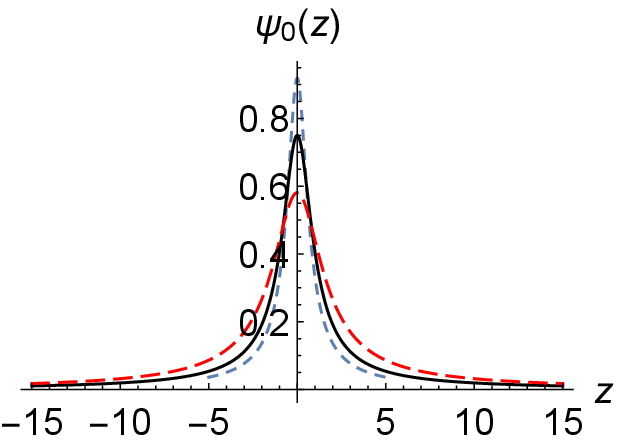}}
\subfigure[$z^2U(z)$]    {\label{z2U}
\includegraphics[width=6cm]{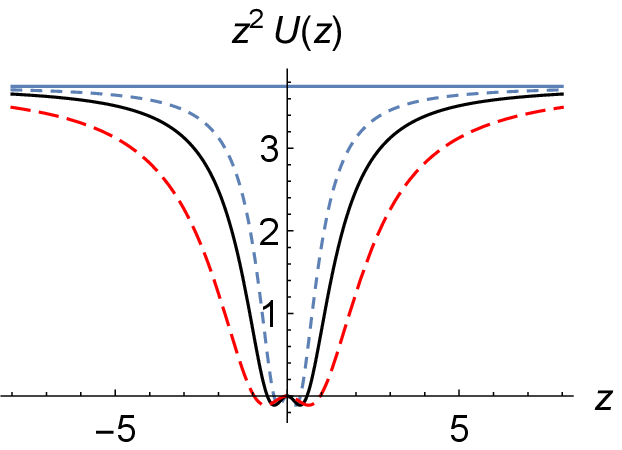}}
\end{center}
\caption{The shapes of the effective potential $U(z)$ and the wave function of the graviton zero mode $\psi_0(z)$ for the second brane solution with $n=2$. The parameters are set to $\kappa=1$, $c=2$ for blue short dashed lines, $c=3$ for black lines, and $c=5$ for red long dashed lines.}
\label{graviton}
\end{figure*}

We now turn our attention to the localization of the graviton zero mode of the second brane solution with $n=2$.
From Eq.~(\ref{solA2}), the integrand of $z=\int \text{e}^{-A(y)}dy$ can be calculated as
\begin{eqnarray}~\label{cr}
z&=&\frac{c}{3 \kappa } \sqrt{\tanh ^2\left(c k^2 y\right)} \text{csch}\left(c k^2 y\right) \text{sech}^{-\frac{3 \kappa }{c^2 k^2}-1}\left(c k^2 y\right) \nonumber\\
&\times&\, _2F_1\left(\frac{1}{2},-\frac{3 \kappa }{2 c^2 k^2};1-\frac{3 \kappa }{2 c^2 k^2};\text{sech}^2\left(c k^2 y\right)\right),
\end{eqnarray}
where $_2F_1$ stands for the hypergeometric function. The above complicated relation between the coordinates $z$ and $y$ makes getting the inverse solution $y(z)$ hopeless. However, if the free parameter $k$ is set to $\sqrt{\frac{3\kappa}{c^2}}$, Eq.~(\ref{cr}) reduces to $z=\frac{c}{3 \kappa } \sinh \left(\frac{3 \kappa  y}{c}\right)$. The inverse solution can be easily obtained as $y=\frac{c}{3 \kappa } \text{arcsinh}\left(\frac{3 \kappa  z}{c}\right)$. Then, the effective potential in the $z$ coordinate is expressed as
\begin{eqnarray}
U(z)=\frac{-66 c^6 \kappa ^2+1107 c^4 \kappa ^4 z^2+16524 c^2 \kappa ^6 z^4+43740 \kappa ^8 z^6}{4 \left(c^4+15 c^2 \kappa ^2 z^2+54 \kappa ^4 z^4\right)^2},
\end{eqnarray}
which is the standard volcano potential (see Fig.~\ref{U}). This potential contains a normalizable graviton zero mode (see Fig.~\ref{psi}):
\begin{eqnarray}
\psi_0(z)=\frac{3 c \sqrt{c^2 \kappa +2 \kappa  \left(c^2+9 \kappa ^2 z^2\right)}}{4 \left(c^2+9 \kappa ^2 z^2\right)^{5/4}},
\end{eqnarray}
where we have inserted the integration constant $C_1=\frac{3}{4}\sqrt{\frac{3\kappa}{c}}$. Except for the localized graviton zero mode, there are a lot of continuous massive Kaluza-Klein (KK) modes, which will lead a correction to the Newtonian potential. As shown in Fig.~\ref{z2U}, $U(z)\sim \frac{15}{4z^2}$ as $|z|\gg 1$, which takes the particular expression $\alpha(\alpha+1)/z^2$. Then, the graviton KK modes on the brane obey the form $\psi_m(0)\sim m^{\alpha-1}$ and the correction for the Newtonian potential between two massive objects at a distance $r$ is $\Delta V(r)\propto 1/r^{2\alpha}$ (see Ref.~\cite{Csaki2000} for more details). For our case $\alpha=3/2$, $|\psi_m(0)|^2\sim m$ for small masses and the correction to the Newtonian potential is $\Delta V(r)\propto 1/r^3$.

\section{Conclusions and Discussions}\label{conclusion}

In this paper, we investigated the thick brane model with a single extra dimension in five-dimensional $f(Q)$ gravity. By adopting the static flat brane metirc and focusing on the particular case with $f(Q)=Q-b Q^n$, we found that the brane system can be solved analytically with the first-order formalism. Then, we investigated two explicit cases with $n=\frac{1}{2}$ and $n=2$, and presented the corresponding thick brane solutions with the first-order formalism. We found that the scalar field for both cases are kink solutions and the energy density always peaks at the origin, which suggest the brane system with no inner structure.

Besides, we investigated the tensor perturbation of the thick brane system. The equation of motion of the tensor perturbation was obtained for a general $f(Q)$. After the Kaluza-Klein (KK) decomposition, this equation can be converted to a Schr$\ddot{\text{o}}$dinger-like equation, and the corresponding Hamiltonian can be factorized as a supersymmetric form, which ensures that there are no tachyonic states. Then, we investigated the localization of the graviton zero modes for both brane systems. Since the wave function of the graviton zero
mode is imaginary and divergent near the origin of the extra dimension for the first brane with $n=\frac{1}{2}$, one can not obtain a localized graviton zero mode in this brane system. For the second brane solution with $n=2$, it was shown that the graviton zero mode can be localized on the brane, which suggests that the four-dimensional Newtonian potential can be recovered on the brane.
What's more, there are a lot of continuous massive KK modes, which may lead corrections to the Newtonian potential. After a brief analysis, we found that the correction is $\Delta V(r)\propto 1/r^3$.
In addition, the stability of the thick brane system under scalar perturbation and the effects of the nonmetricity on the scalar perturbation are also interesting problems. These are left for our future works.

\acknowledgments{
This work was supported by the National Natural Science Foundation of China (No. 11875151), and Scientific Research Program Funded by Shaanxi Provincial Education Department (No. 20JK0553).
}


\begin{thebibliography}{00}

\bibitem{Nester1999}
J. M. Nester and H. Yo,
Chin. J. Phys. {\bf 37}, 113 (1999).

\bibitem{Jimenez2018a}
J. B. Jim$\acute{\text{e}}$nez, L. Heisenberg, and T. Koivisto,
Phys. Rev. D {\bf 98}, 044048 (2018).



\bibitem{Lazkoz2019}
R. Lazkoz, F. S. N. Lobo, M. Ortiz-Ba$\tilde{\text{n}}$os, and V. Salzano,
Phys. Rev. D {\bf 100}, 104027 (2019).

\bibitem{Mandal2020a} 	
S. Mandal, P. K. Sahoo, and J. R. L. Santos,
Phys. Rev. D {\bf 102}, 024057 (2020).

\bibitem{Jianbo2019}
J. B. Lu, X. Zhao, and G. Y. Chee,
Eur. Phys. J. C {\bf 79}, 530 (2019).

\bibitem{Hohmann2019}
M. Hohmann, C. Pfeifer, J. L. Said, and U. Ualikhanova,
Phys. Rev. D {\bf 99}, 024009 (2019).

\bibitem{Soudi2019} 	
I. Soudi, G. Farrugia, V. Gakis, J. L. Said, and E. N. Saridakis,
Phys. Rev. D {\bf 100}, 044008 (2019).

\bibitem{Jarv2018}
L. J$\ddot{\text{a}}$rv, M. R$\ddot{\text{u}}$nkla, M. Saal, and O. Vilson,
Phys. Rev. D {\bf 97}, 124025 (2018).

\bibitem{Runkla2018}
M. R$\ddot{\text{u}}$nkla and O. Vilson,
Phys. Rev. D {\bf 98}, 084034 (2018).

\bibitem{Harko2018}
T. Harko, T. S. Koivisto, F. S. N. Lobo, G. J. Olmo, and D. Rubiera-Garcia,
Phys. Rev. D {\bf 98}, 084043 (2018).

\bibitem{Yixin2019} 	
Y. X. Xu, G. J. Li, T. Harko, and S. D. Liang,
Eur. Phys. J. C {\bf 79}, 708 (2019).

\bibitem{Jimenez2018b}
J. B. Jim$\acute{\text{e}}$nez, L. Heisenberg, and T. S. Koivisto,
JCAP {\bf 1808}, 039 (2018).

\bibitem{Jimenez2019}
J. B. Jim$\acute{\text{e}}$nez, L. Heisenberg, and T. S. Koivisto,
Universe {\bf 5}, 173 (2019).

\bibitem{Dialektopoulos2019}
K. F. Dialektopoulos, T. S. Koivisto, and S. Capozziello,
Eur. Phys. J. C {\bf 79}, 606 (2019).

\bibitem{Jimenez2020} 	
J. B. Jim$\acute{\text{e}}$nez, L. Heisenberg, T. S. Koivisto, and S. Pekar,
Phys. Rev. D {\bf 101}, 103507 (2020).

\bibitem{Mandal2020b}
S. Mandal, D. Wang, and P. K. Sahoo,
Phys. Rev. D {\bf 102}, 124029 (2020).






\bibitem{rs1}
L.  Randall  and  R.  Sundrum,  Phys.  Rev.  Lett.  {\bf 83},  3370 (1999).

\bibitem{rs2}
L.  Randall  and  R.  Sundrum,  Phys.  Rev.  Lett.  {\bf 83},  4690 (1999).

\bibitem{Davoudiasl2000}
H. Davoudiasl, J. L. Hewett, and T. G. Rizzo, Phys. Lett. B {\bf 473}, 43 (2000).

\bibitem{Gherghetta2001}
T. Gherghetta and A. Pomarol, Nucl. Phys. B {\bf 602}, 3 (2001).

\bibitem{Huber2001}
S. J. Huber and Q. Shafi, Phys. Lett. B {\bf 498}, 256 (2001).





\bibitem{DeWolfe2000}
O. DeWolfe, D. Z. Freedman, S. S. Gubser, and A. Karch, Phys. Rev. D {\bf 62}, 046008 (2000).

\bibitem{Csaki2000}
C. Csaki, J. Erlich, T. J. Hollowood, and Y. Shirman, Nucl. Phys. B {\bf 581}, 309 (2000).

\bibitem{Gremm2000}
M. Gremm, Phys. Lett. B {\bf 478}, 434 (2000).

\bibitem{Giovannini2001}
M. Giovannini, Phys. Rev. D {\bf 64}, 124004 (2001).

\bibitem{Minamitsuji2006}
M. Minamitsuji, W. Naylor, and M. Sasaki, Phys. Lett. B {\bf 633}, 607 (2006).

\bibitem{Dzhunushaliev2008}
V. Dzhunushaliev, V. Folomeev, D. Singleton, and S. Aguilar-Rudametkin, Phys. Rev. D {\bf 77}, 044006 (2008).

\bibitem{Liu2011}
Y. X.  Liu,  Y.  Zhong,  Z. H.  Zhao,  and  H. T.  Li,  J. High Energy Phys. {\bf 1106}, 135 (2011).

\bibitem{Zhong2011}
Y. Zhong, Y. X. Liu, and K. Yang, Phys. Lett. B {\bf 699}, 398 (2011).

\bibitem{Bazeia2015}
D. Bazeia, A. S. Lobao, and R. Menezes, Phys. Lett. B {\bf 743}, 98 (2015).

\bibitem{Dzhunushaliev2020}
V. Dzhunushaliev, V. Folomeev, G. Nurtayeva, and S. D. Odintsov,
Int. J. Geom. Meth. Mod. Phys. {\bf 17}, 2050036 (2020).

\bibitem{Bazeia2020a}
D. Bazeia, D. A. Ferreira, and D. C. Moreira,
EPL {\bf 129}, 11004 (2020).

\bibitem{Bazeia2020b}
D. Bazeia, D. A. Ferreira, and M. A. Marques,
Eur. Phys. J. Plus {\bf 135}, 587 (2020).

\bibitem{Sui2020}
T. T. Sui, W. D. Guo, Q. Y. Xie, and Y. X. Liu,
Phys. Rev. D {\bf 101}, 055031 (2020).

\bibitem{Cui2020}
Z. Q. Cui, Z. C. Lin, J. J. Wan, Y. X. Liu, and L. Zhao,
J. High Energy Phys. {\bf 2012}, 130 (2020).

\bibitem{Moreira2101}
A. R. P. Moreira, J. E. G. Silva, F. C. E. Lima, and C. A. S. Almeida,
arXiv:2101.10054.

\bibitem{Melfo2008}
A. Melfo, N. Pantoja, and J. D. Tempo, Phys. Rev. D {\bf 73}, 044033 (2006).

\bibitem{Liu2008}
Y. X.  Liu,  X. H.  Zhang,  L. D.  Zhang,  and  Y. S.  Duan, J. High Energy Phys. {\bf 0802}, 067 (2008).

\bibitem{Liu2009}
Y. X. Liu, J. Yang, Z. H. Zhao, C. E. Fu, and Y. S. Duan, Phys. Rev. D {\bf 80}, 065019 (2009).

\bibitem{Dantas2015}
D. M.  Dantas, D. F. S.  Veras,  J. E. G.  Silva,  and  C. A. S. Almeida, Phys. Rev. D {\bf 92}, 104007 (2015).

\bibitem{Vaquera-Araujo2015}
C. A. Vaquera-Araujo and O. Corradini, Eur. Phys. J. C {\bf 75}, 48 (2015).

\bibitem{Arai2013}
M. Arai, F. Blaschke, M. Eto, and N. Sakai, J. Phys. Conf. Ser. {\bf 411}, 012001 (2013).

\bibitem{Zhao2015}
Z. H. Zhao, Q. Y. Xie, and Y. Zhong, Class. Quant. Grav. {\bf 32}, 035020 (2015).



\bibitem{Kobayashi2002}
S. Kobayashi, K. Koyama, and J. Soda, Phys. Rev. D {\bf 65}, 064014 (2002).

\bibitem{Andrianov2008}
A. A. Andrianov and L. Vecchi, Phys. Rev. D {\bf 77}, 044035 (2008).

\bibitem{Barbosa-Cendejas2008}
N. Barbosa-Cendejas, A. Herrera-Aguilar, M. A. R. Santos, and C. Schubert, Phys. Rev. D {\bf 77}, 126013 (2008).

\bibitem{Andrianov2013}
A. A. Andrianov, V. A. Andrianov, and O. O. Novikov, Eur. Phys. J. C {\bf 73}, 2675 (2013).

\bibitem{Barbosa-Cendejas2005}
N.  Barbosa-Cendejas  and  A.  Herrera-Aguilar,  J. High Energy Phys. {\bf 0510}, 101 (2005).

\bibitem{Veras2016}
D. F. S. Veras, W. T. Cruz, R. V. Maluf, and C. A. S. Almeida, Phys. Lett. B {\bf 754}, 201 (2016).









\bibitem{Liu2017}
Y. X. Liu, arXiv:1707.08541.




\end{thebibliography}
\end{document}